 \documentclass[final,3p,times,twocolumn]{elsarticle}
\usepackage{amssymb}
\usepackage{graphicx}% Include figure files
\usepackage{epstopdf}
\usepackage[caption=false]{subfig}
\def \azno{a-ZnO}
\def \izo{a-IZO}

\journal{Journal of Non-Crystalline Solids}

\begin{document}

\begin{frontmatter}



\title{{\it Ab initio} model of amorphous zinc oxide ({\azno}) and  a-X$_{0.375}$Z$_{0.625}$O (X= Al, Ga and In)}


\author{Anup Pandey}

\address{Department of Physics and Astronomy, Condensed Matter and Surface Science Program, Ohio University, Athens OH 45701, USA}

\author{Heath Scherich}
\address{Department of Physics and Astronomy, Ohio University, Athens OH 45701, USA}

\author{D. A. Drabold}
\address{Department of Physics and Astronomy, Ohio University, Athens OH 45701, USA}

\begin{abstract}
Density functional theory (DFT) calculations are carried out to study the structure and electronic structure of amorphous zinc oxide (\azno). The models were prepared by the "melt-quench" method. The models are chemically ordered  with some coordination defects. The effect of trivalent dopants in the structure and electronic properties of {\azno} is investigated. Models of a-X$_{0.375}$Z$_{0.625}$O (X= Al, Ga and In) were also prepared by the "melt-quench" method. The trivalent dopants reduce the four-fold Zn and O, thereby introducing some coordination defects in the network. The dopants prefer to bond with O atom. The network topology is discussed in detail. Dopants reduce the gap in EDOS by producing defect states minimum while maintaining the extended nature of the conduction band edge. 

\end{abstract}

\begin{keyword}


amorphous Zinc Oxide, doping, EDOS, DFT 

\end{keyword}

\end{frontmatter}


%\section{Anup}
%\label{pandey}
\section{Introduction}
Crystalline ZnO has important application as a piezoelectric material and due to its property of being transparent in visible light \cite{kordesch}. It has a wide direct band gap ($\sim$ 3.37 eV at 300 K) which makes it a promising candidate for optoelectronic devices ~\cite{kordesch,huang}. Therefore, there has been a wealth of  experimental work in crystalline ZnO ({\azno}). On the other hand, the study of amorphous ZnO is still in its nascent stage compared to its crystalline counterpart as far as the experimental work is concerned. 

The amorphous transparent oxide materials have immense use in device technology~\cite{nature}. Ionic amorphous oxide semiconductors like {\azno} have high electron mobility ($\sim$ 5-40 cm$^{2}$/V s) compared to the covalent amorphous semiconductor like a-Si ($\sim$ 1 cm$^{2}$/V s)which make them a better candidate for the device application such as thin film transistors (TFTs)~\cite{robertson}. Experimentally, various techniques such as pulse laser deposition ~\cite{pld}, molecular beam epitaxy~\cite{mbe}, radio-frequency magnetron sputtering~\cite{rf} etc. have been used to make {\azno}   and the structure obtained is highly dependent on the substrate material and temperature.   
There are advantages of {\azno} over its crystalline counterpart. First, it is easier and cost efficient to produce a large amorphous sheet compared to a large single crystal. Also, the {\azno} has been prepared at low temperature ($\sim$ 300 K) compared to crystalline ZnO ($\sim$ 800 K - 1100 K) and its optical properties over the wide spectral range has been investigated~\cite{kordesch}. On doping trivalent elements such as  Al, Ga and In  on {\azno} mobility can be increased significantly ~\cite{hosono}.

In this work, we report the structure and electronic properties of amorphous phases of ZnO and {\azno} doped with trivalent dopant atoms such as Indium (In), Gallium (Ga) and Aluminium (Al) using a plane wave basis density functional theory (DFT) and comparisons with the experiments and other molecular dynamics (MD) simulations are made when possible. For the first time, accurate methods are used to compute the topological and chemical order of the materials and determine electronic characteristics.

The paper is arranged as follows. In Section II, we provide details about the computational technique used in modeling various models. In Section III, we present the simulation results for {\azno} and a-X$_{0.375}$Z$_{0.625}$O (X= Al, Ga and In) and make comparisons with experiments and other MD methods.

\begin{figure}
%\begin{center}
\includegraphics[width=3.0 in]{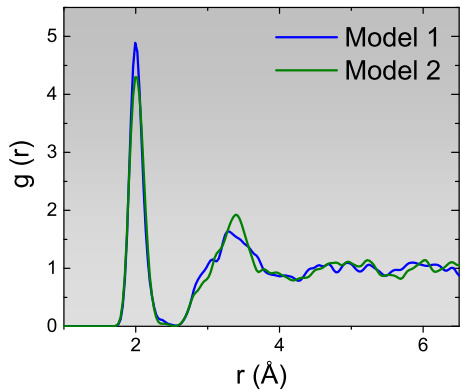}
\hfill
\caption{
(color online) The total radial distribution function (RDF) for four {\azno} models. Model 1 and Model 2 corresponds to the models obtained by two different quenching rates as described in the method section. Blue is for Model 1 and green is for Model 2. 
}
\label{znordfs}
%\end{center}
\end{figure}

\section{Computational Methods} 
{\it Ab initio} calculations are performed using the DFT code VASP~\cite{vasp1,vasp2,vasp3} using projected augmented plane waves (PAW)~\cite{paw} with Perdew-Burke-Ernzerhof (PBE) exchange-correlation functional~\cite{pbe} and a plane-wave cutoff off energy of 300 eV. All calculations were carried out at $\Gamma$ point. We have prepared a model by using the melt-quench method. For {\azno}, the system consists of 128 atoms in a cubic box of length 12.34 {\AA} corresponding to the experimental density of 4.6 g/cm$^{3}$~\cite{huang}. The random starting configuration is equilibrated at 5000 K is cooled to 3000 K at 100 K/ps followed by an equilibration of 5 ps. The structure at 3000 K is cooled in steps to temperatures 2300 K, 1600 K and 300 K at the rate of 50 K/ps followed by 5 ps equilibration in each temperature. Finally, the structure at 300 K is quenched to 0 K at the rate of 25 K/ps which is again followed by equilibration of 5 ps. The structure is then relaxed using the conjugate gradient (CG) method. This model is termed as Model 1. To contrast different quench rates, the configuration at 3000 K was also cooled to 300 K at a rate of 180 K/ps followed by the equilibration of 5 ps. The model is relaxed using CG method. Finally, the equilibrated structure is quenched to 0 K at a rate of 50 K/ps and then equilibrated for another 5 ps. We call this model as Model 2. 

For a-X$_{0.375}$Z$_{0.625}$O (X= Al, Ga and In), a random starting configuration of 128 atoms was melted at 5000 K followed by cooling to 3000 K at 100 K/ps and then equilibrated for 5 ps. A schedule of cooling is carried out at temperatures 2300 K, 1600 K and 300 K at a rate of 25 K/ps followed by the equilibration of 5 ps in each temperature. Finally, the structures are cooled to 0 K at the rate of 40 K/ps followed by the equilibration of 5 ps. The final structures are volume relaxed to estimate the density. The cubical box lengths for Al-, Ga- and In-doped models after volume relaxations are 12.26 {\AA}, 12.28 {\AA} and 12.31 {\AA} respectively. We find no strong dependence on the quenching processes.
%

%\begin{figure}
%\subfloat[\label{znopartials0k}]{%
%\includegraphics[width=3.5 in]{ZnO_partials.eps}%
% }\hfill
%\subfloat[\label{znopartials300k}]{%
%\includegraphics[width=3.5 in]{ZnO_300K_partials.eps}
%}
%\caption{(color online) Partial pair correlation functions for 128-atom model {\azno} obtained by slow and %fast quenching at (a)0 K and (b) 300 K. Blue is for fast and red is for slow quenching rate.}
%\label{znopartials}
%\end{figure}

%
\begin{figure}
\begin{center}
\includegraphics[width=3.8 in]{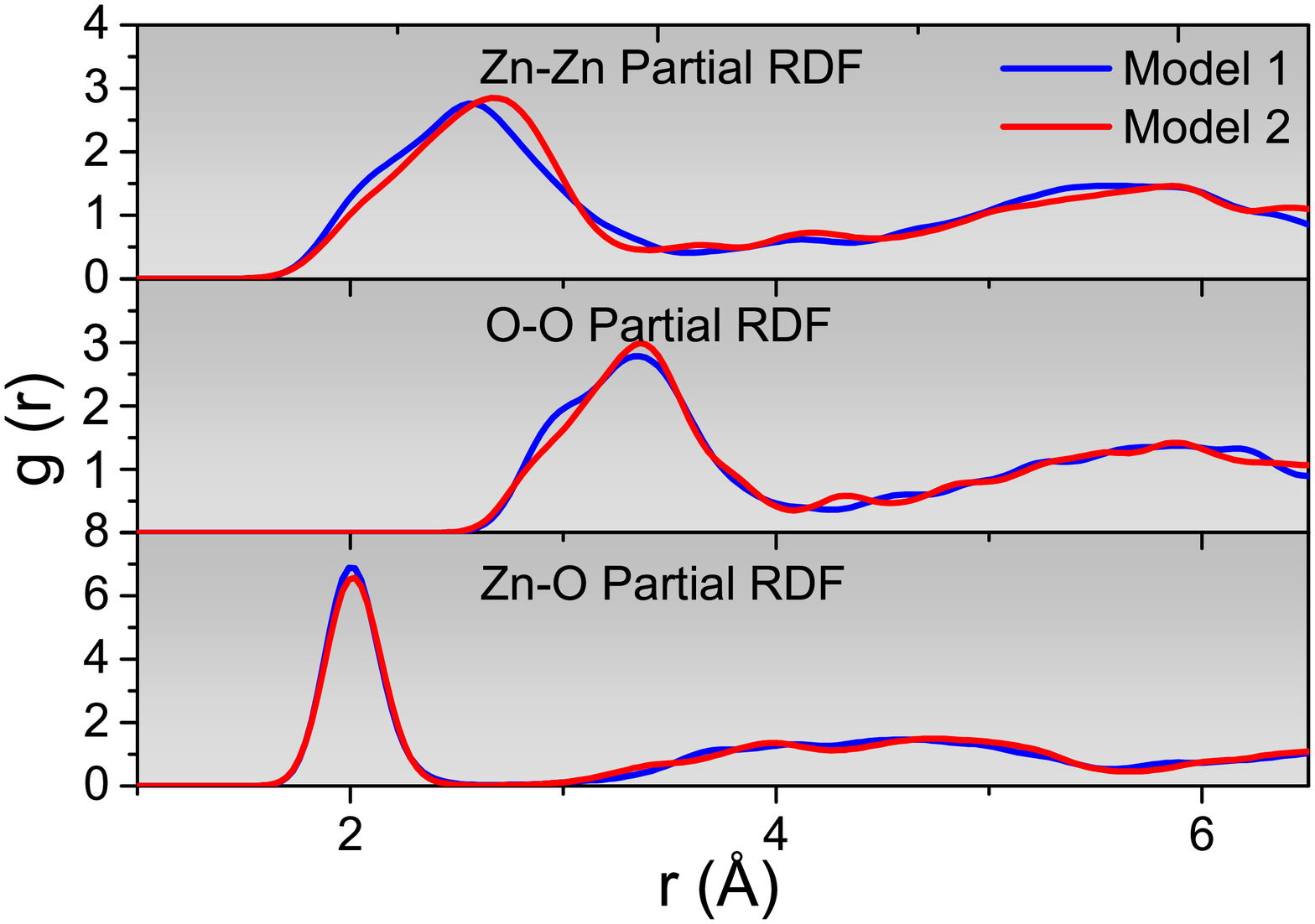}
\hfill
\caption{
(color online) Partial pair correlation functions for 128-atom model {\azno}. Blue is for Model 1 and red is for Model 2 as described in the method section.
}
\label{znopartials300k}
\end{center}
\end{figure}
\section{Results}
\subsection{Amorphous Zinc Oxide ({\azno})} 
 Structural properties are investigated by the  radial distribution functions (RDFs) and partial radial distribution functions. The total RDFs for Model 1 and Model 2 are shown in Fig.\ref{znordfs}. The partial pair correlation functions for Zn-Zn, Zn-O and O-O are shown in Fig.\ref{znopartials300k}. The partials for both models show similar features. For both Zn-Zn and O-O partials, the first peak is around 3.40 {\AA} while for Zn-O the first peak position is at 2.00 {\AA} as shown in Table \ref{tab2}. 
\begin{table}
\caption{ 
The coordination number for Zn and O expressed in percentage, average coordination number and the DFT-GGA energy for {\azno} model. The coordination number for Zn are compared with the other MD model~\cite{huang}.
As expected there are a few more coordination defects in the more rapidly quenched model 2.
}
 \begin{center}
\begin{tabular}[b]{|p {1.8cm}|p {1.4cm}|p {1.2cm}|p {1.2cm}|}

  \hline
   %\multicolumn{4}{|c|}{Peak position ($\AA$)} \\ \hline
    & Model 1 & Model 2 & MD (Ref.(\cite{huang})\\ \hline
   Zn-Zn (\%)  & 0 &0 &-\\ \hline
     O-O (\%)& 0 &0 &-\\ \hline
      Zn$_{3}$(\%)& 15.63 &23.44&32.00\\ \hline
      Zn$_{4}$(\%)& 81.25 &75.00&60.00\\ \hline
      Zn$_{5}$(\%) & 3.12 &1.56&7.00\\ \hline
      O$_{3}$(\%) & 15.63  &25.00&-\\ \hline
      O$_{4}$(\%) & 81.25 &71.88&-\\ \hline
      O$_{5}$(\%) & 3.12 &3.12&-\\ \hline
      \emph{n}$_{Zn}$ &3.88 &3.78&-\\ \hline
      \emph{n}$_{O}$ & 3.88 &3.78&-\\ \hline
      Energy (eV/atom) &-4.36&-4.34&-\\ \hline
      \hline
\end{tabular}
\label{tab1}
\end{center}
\end{table}

\begin{figure}
\begin{center}
\includegraphics[width=3.2 in]{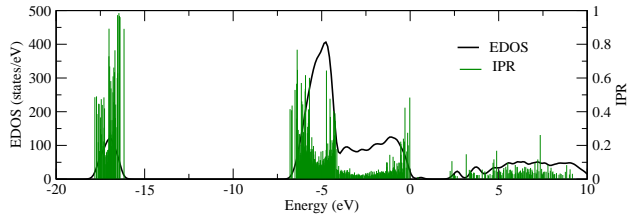}
\hfill
\caption{
(color online) (black) Electronic density of states of the 128-atom model {\azno} at 300 K (Model I) obtained using GGA-PBE density functional theory calculation. The green vertical lines represent the inverse participation ratio (IPR) used to measure the electronic state localization. Longer IPR implies strong localization. The Fermi level is at 0.28 eV. The PBE gap is 1.36 eV. 
}
\label{edos300ks}
\end{center}
\end{figure}
\begin{figure}
\subfloat[\label{izordf}]{%
\includegraphics[width=3.5 in]{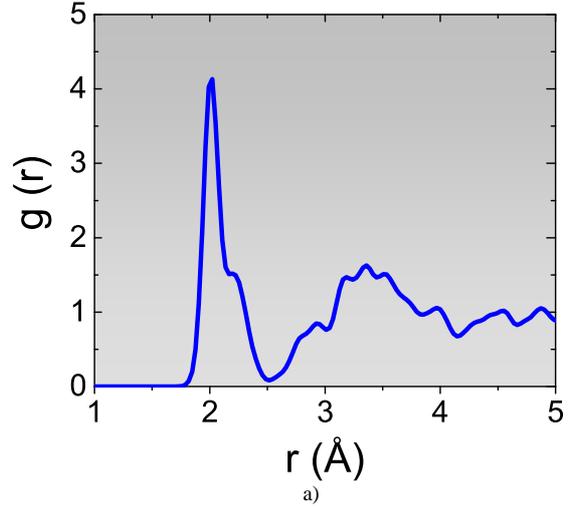}%
 }\hfill
 \subfloat[\label{izopartials}]{%
\includegraphics[width=3.8 in]{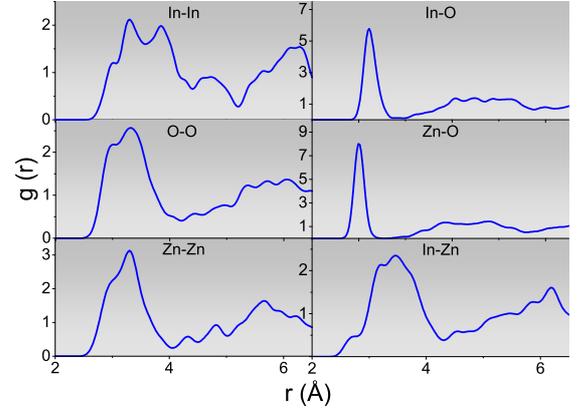}
}
\caption{(color online)(a) Total pair correlation functions for 128-atom model {\izo} at 300 K. (b) Partial pair correlation function of {\izo} at 300 K.}
\label{izoall}
\end{figure}
\begin{figure}
\begin{center}
\includegraphics[width=3.2 in]{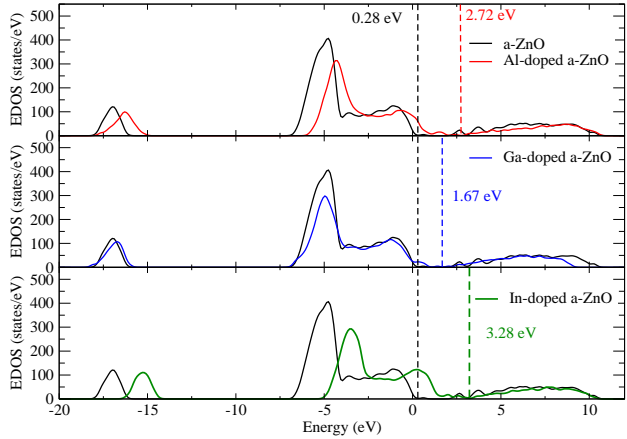}
\hfill
\caption{
(color online) Electronic density of states for a-X$_{0.375}$Z$_{0.625}$O (X= Al, Ga and In) models compared to that of {\azno} at 300 K. The Fermi levels are shown by vertical broken lines. 
}
\label{edosxall}
\end{center}
\end{figure}
The network is chemically ordered. We calculated the coordination number for Zn and O given by the average first neighbour atoms around Zn or O as a reference atom. Most of the atoms are four-fold coordinated with above 75\% of four-fold coordinated Zn and O in all four models. Our models exhibit a higher percentage of four-fold Zn  compared to the model obtained from empirical molecular dynamics simulation model ($\sim$ 60\%)~\cite{huang}. The average coordination number for Zn (\emph{n}$_{Zn}$) and O (\emph{n}$_{O}$) and the 3-, 4- and 5-fold coordinated Zn and O, denoted by the respective subscript (Zn$_{3}$, O$_{3}$,...) is shown in Table \ref{tab1}. The DFT energy per atom for Model 1 is -4.36 eV/atom and for Model 2 is -4.34 eV/atom (Table \ref{tab1}). This suggest that Model 1 is energetically favorable compared to Model 2.

The electronic structure is analysed by calculating the electronic density of states (EDOS) and inverse participation ratio (IPR) of the individual states. The EDOS is shown in Fig.\ref{edos300ks} (black) and the green vertical lines represents the electronic state localization measured by IPR~\cite{raymond,kiran}. The value of IPR is 1 for highly localized state and 1/N for extended state, where N is the number of atoms. The IPR in Fig.\ref{edos300ks} shows that the localization of valence tail states much larger than the conduction tail states. Thus, the mobility of \emph{n}-type of carrier is expected to be much higher than the \emph{p}-type. This feature supports the  asymmetry in the localization of valence and conduction band tail states in amorphous metal oxide by Robertson~\cite{robertson}. Similar asymmetrical behavior in amorphous gallium nitride was shown by Cai and Drabold~\cite{dad_gan}.

The band gap (gap between the highest occupied electronic state and the lowest unoccupied electronic state), in the model is 1.36 eV which is slightly less than the experimental band gap of 1.60 eV between the valence band edge and Zn4s4p states~\cite{dieter}. The underestimate of the band gap is of course expected in the DFT-GGA calculation. 

\subsection{ Al-, Ga- and In-doped amorphous ZnO: a-X$_{0.375}$Z$_{0.625}$O (X= Al, Ga and In)}
To investigate the effect of trivalent dopants on local coordination and electronic structure of {\azno}, 37.5{\%} of Zn is replaced by group III elements X (X= Al, Ga and In) to model a-X$_{0.375}$Z$_{0.625}$O. The  atomic percentage of dopants in all the models is 18.75\%. The effect of dopants in the structure and electronic properties are investigated by RDFs, partial pair correlation functions and electronic density of states.

\begin{table}
\caption{ 
First peak position for Zn-Zn, Zn-O and O-O partial pair correlation functions of {\azno} (Model 1) and a-X$_{0.375}$Z$_{0.625}$O (X= Al, Ga and In) models.
}
 \begin{center}
\begin{tabular}[b]{|p {1.0cm}|p {0.8cm}|p {2.0cm}|p {2.0cm}|p {2.0cm}|}

  \hline
   \multicolumn{5}{|c|}{Peak position for 300 K (slow) models ({\AA})} \\ \hline
    First peak &ZnO  & Al$_{0.375}$Z$_{0.625}$O & Ga$_{0.375}$Z$_{0.625}$O& In$_{0.375}$Z$_{0.625}$O\\ \hline
   Zn-Zn & 3.40  & 2.46 & 2.90 & 3.20 \\ \hline
     O-O & 2.00  & 1.90 & 2.00 & 2.00 \\ \hline
     O-O & 3.40  & 3.00 & 3.10 & 3.20 \\ \hline
      \hline
\end{tabular}
\label{tab2}
\end{center}
\end{table}
The total RDFs for In-doped {\azno} is shown in Fig.\ref{izordf}.
The Zn-O correlation is not affected by the presence of dopants while there is a slight decrease in the correlation peaks for Zn-Zn and O-O which is illustrated in Table \ref{tab2}. The peak positions are obtained from the partial pair correlation functions shown in Fig.\ref{izopartials} for In-doped {\azno} and similar plots for Al-, and Ga-doped {\azno} which is not shown here. The 4-fold Zn and O are reduced significantly in all three doped models. The Al and In bond only with O. In Al-doped model, 95.83\% and 4.17\%  Al forms 4-fold and 3-fold bond with O. In In-doped model, 20.83\%,  25.00\%, 8.33\%, 37.5\% and 8.34\% In form 6-, 5-, 4-, 3- and 2-fold bond with O atom. In Ga-doped model, 58.33\% and 33.33\% Ga form 4-fold and 3-fold bond with O while 8.34\% Ga form 3-fold with O and 1-fold with Zn. This suggest that the group III elements are more likely to form a bond with O while doped in {\azno}.

The Zn-Zn and In-In distances in our model are around 3.20 {\AA} and 3.50 {\AA} which is close to 3.20-3.40 {\AA} for Zn-Zn and 3.30-3.6 {\AA} for In-In of classical MD model~\cite{ramo}. This compares well with the average metal-metal peak in x-ray diffraction measurements of IZO thin layers~\cite{eguchi}. Also, the Zn-O and In-O distances in our model are around 2.00 {\AA} and 2.20 {\AA} compared to the 1.95 {\AA| and 2.20 {\AA} respectively of the classical MD model~\cite{ramo}. These peak positions are consistent with the metal-oxygen peaks at 2.12-2-14 {\AA} in the experiment~\cite{eguchi}.

The electronic density of states (EDOS) for a-X$_{0.375}$Z$_{0.625}$O (X= Al, Ga and In) models compared to that of {\azno} at 0 K is shown in Fig.\ref{edosxall}. The introduction of dopants clearly reduces the gap by introducing the defect states. The localized states near the valence band edge induced by doping can be inferred to the increase in undercoordinated O atoms in the network introduced by doping. The conduction band edge is unaltered by the addition of dopant elements. The extended nature of the conduction band is preserved by the addition of dopants which is in accordance to the conclusion by Hosono~\cite{hosono}. On the other hand, in crystalline ZnO doped by group III elements Al, Ga and In, the dopants form extra localized level in the conduction band, which modifies the conduction band and reduces the optical band gap~\cite{sans}.

\section{Conclusions}
In conclusion, we have created models of amorphous zinc oxide ({\azno}) using melt-quench method and studied their structural and electronic properties in detail. The electronic band gap of our model is 1.36 eV which is in reasonable agreement  with the experimental band gap 1.60 eV. We have calculated the DFT energies for the two models of {\azno} obtained by different quenching rate for comparison. The effect of trivalent dopants in the local structure and the electronic structure of {\azno} is investigated in detail by preparing a-X$_{0.375}$Z$_{0.625}$O (X= Al, Ga and In) models by melt-quench method. The dopants reduce the number of 4-fold coordinated Zn and O in the network and most of them prefer to bond with oxygen. The electronic gap is reduced by the presence of defect states by forming undercoordinated O states in the valence band edge while the conduction band edge is still extended. Coordinates are available upon request.

\section{Acknowledgements}
We thank the US NSF under the grants DMR 150683, 1507166 and 1507670 for supporting 
this work, and the Ohio Supercomputer Center for computer time. We acknowledge the financial support of Condensed Matter and Surface Science program of Ohio University.



\begin{thebibliography}{00}

\bibitem{kordesch}Khoshman, J.M. and Kordesch, M.E., 2007. Optical constants and band edge of amorphous zinc oxide thin films. Thin Solid Films, 515(18), pp.7393-7399.

\bibitem{huang}Lin, K.H., Sun, S.J., Ju, S.P., Tsai, J.Y., Chen, H.T. and Hsieh, J.Y., 2013. Observation of the amorphous zinc oxide recrystalline process by molecular dynamics simulation. Journal of applied physics, 113(7), p.073512.

\bibitem{nature}Nomura, K., Ohta, H., Takagi, A., Kamiya, T., Hirano, M. and Hosono, H., 2004. Room-temperature fabrication of transparent flexible thin-film transistors using amorphous oxide semiconductors. Nature, 432(7016), pp.488-492.

\bibitem{robertson}Robertson, J., 2008. Physics of amorphous conducting oxides. Journal of Non-Crystalline Solids, 354(19), pp.2791-2795.

\bibitem{pld}Hayamizu, S., Tabata, H., Tanaka, H. and Kawai, T., 1996. Preparation of crystallized zinc oxide films on amorphous glass substrates by pulsed laser deposition. Journal of applied physics, 80(2), pp.787-791.

\bibitem{mbe}Bagnall, D.M., Chen, Y.F., Zhu, Z., Yao, T., Koyama, S., Shen, M.Y. and Goto, T., 1997. Optically pumped lasing of ZnO at room temperature. Applied Physics Letters, 70(17), pp.2230-2232.

\bibitem{rf}Xingwen, Z., Yongqiang, L., Ye, L., Yingwei, L. and Yiben, X., 2006. Study on ZnO thin films deposited on sol–gel grown ZnO buffer by RF magnetron sputtering. Vacuum, 81(4), pp.502-506.

%\bibitem{mott}Davis, E.A. and Mott, N., 1970. Conduction in non-crystalline systems V. Conductivity, optical %absorption and photoconductivity in amorphous semiconductors. Philosophical Magazine, 22(179), pp.0903-0922.

\bibitem{hosono}Hosono, H., 2006. Ionic amorphous oxide semiconductors: Material design, carrier transport, and device application. Journal of Non-Crystalline Solids, 352(9), pp.851-858.

\bibitem{sans}Sans, J.A., Sánchez-Royo, J.F., Segura, A., Tobias, G. and Canadell, E., 2009. Chemical effects on the optical band-gap of heavily doped ZnO: M III (M= Al, Ga, In): An investigation by means of photoelectron spectroscopy, optical measurements under pressure, and band structure calculations. Physical Review B, 79(19), p.195105.

\bibitem{vasp1}Kresse, G. and Hafner, J., 1993. Ab initio molecular dynamics for liquid metals. Physical Review B, 47(1), p.558.

\bibitem{vasp2} Kresse, G. and Furthmüller, J., 1996. Efficient iterative schemes for ab initio total-energy calculations using a plane-wave basis set. Physical review B, 54(16), p.11169.

\bibitem{vasp3} Kresse, G. and Furthmüller, J., 1996. Efficiency of ab-initio total energy calculations for metals and semiconductors using a plane-wave basis set. Computational Materials Science, 6(1), pp.15-50.

\bibitem{paw}Kresse, G. and Joubert, D., 1999. From ultrasoft pseudopotentials to the projector augmented-wave method. Physical Review B, 59(3), p.1758.

\bibitem{pbe}Perdew, J.P., Burke, K. and Ernzerhof, M., 1996. Generalized gradient approximation made simple. Physical review letters, 77(18), p.3865.

\bibitem{raymond}Atta-Fynn, R., Biswas, P. and Drabold, D.A., 2004. Electron–phonon coupling is large for localized states. Physical Review B, 69(24), p.245204.

\bibitem{kiran}Prasai, K., Biswas, P. and Drabold, D.A., 2016. Electrons and phonons in amorphous semiconductors. Semiconductor Science and Technology, 31(7), pp.73002-73015.

\bibitem{dad_gan}Stumm, P. and Drabold, D.A., 1997. Can amorphous GaN serve as a useful electronic material?. Physical review letters, 79(4), p.677.

\bibitem{dieter}Schmeißer, D., Haeberle, J., Barquinha, P., Gaspar, D., Pereira, L., Martins, R. and Fortunato, E., 2014. Electronic structure of amorphous ZnO films. physica status solidi (c), 11(9‐10), pp.1476-1480.

\bibitem{ramo}Ramo, D.M., Chroneos, A., Rushton, M.J.D. and Bristowe, P.D., 2014. Effect of trivalent dopants on local coordination and electronic structure in crystalline and amorphous ZnO. Thin Solid Films, 555, pp.117-121.

\bibitem{eguchi}Eguchi, T., Inoue, H., Masuno, A., Kita, K. and Utsuno, F., 2010. Oxygen Close-Packed Structure in Amorphous Indium Zinc Oxide Thin Films. Inorganic chemistry, 49(18), pp.8298-8304.

\end{thebibliography}


\end{document}